\documentclass[11pt,onecolumn]{IEEEtran}

\usepackage{amsmath,url}
\usepackage{epsfig,amssymb,amsbsy,verbatim,array,enumerate}
\usepackage{pstricks,psfrag,theorem,cite,paralist}

\let\intern=\iftrue

\def\figref#1{Fig.\,\ref{#1}}%
\def\E{\mathbb{E}}
\def\P{\mathbb{P}}
\def\R{\mathbb{R}}

\def\ie{{\em i.e.}}

\def\dd{\mathrm{d}}

\def\calV{\mathcal{V}}
\def\calN{\mathcal{N}}
\def\calP{\mathcal{P}}

\def\ppp{\Phi_{\rm PPP}}
\def\lppp{\lambda_{\rm PPP}}
\def\gbs{g_{\rm BS}}
\def\cc{\gamma}

\newtheorem{lemma}{Lemma}

\newtheorem{definition}{Definition}

\newlength{\figwidth}


\begin{document}
\title{User Point Processes in Cellular Networks} 
\author{Martin Haenggi\\Dept.~of Electrical Engineering\\University of Notre Dame 
\thanks{Manuscript date \today. The support of the U.S.~NSF (grant CCF 1525904) is gratefully acknowledged.}
}
\maketitle
\begin{abstract}
The point process of concurrent users is critical for the analysis of cellular networks,
in particular for the uplink and for full-duplex communication. We analyze the properties
of two popular models.
For the first one, we provide an accurate
characterization of the pair correlation functions from the user and the base station
point of view, which are applied to approximate the user process by Poisson and
Ginibre point processes.
For the second model, which includes the first model asymptotically,
we study the cell vacancy probability, the mean area of vacant and occupied cells,
 the user-base station distance, and the
pair correlation function in lightly and heavily loaded regimes.
\end{abstract}
\section{Introduction}
{\em Motivation.}
Our goal is to analyze the point process of users who are served in the
same time-frequency resource block (RB) in a cellular network.
These users are those who interfere with each other in the uplink and when
using full-duplex transmission or co-channel D2D communication.

\medskip
{\em Prior work.}
There are two popular  point process models for the users, which we call
models of type I and II. However, no results are available for
their pair correlation function (pcf) $g(r)$, which is critical for approximations by simpler, more
tractable point processes.
For the pcf between the
typical base station (BS) and the users in a Poisson point process (PPP) of BSs
of intensity $\lambda$, the approximation $\gbs(r)\approx 1-e^{-\lambda \pi r^2}$ is suggested
in \cite[Remark 1]{net:Singh15twc}. However, this approximation is only accurate for lightly
loaded networks, where a significant fraction of BSs are idle in the RB considered.
We provide tight approximations for $g(r)$ and $\gbs(r)$ for heavily and lightly loaded networks.
Accordingly, PPPs with intensity functions
$\lambda g(r)$ or $\lambda \gbs(r)$, respectively, provide accurate approximations for the
user point processes. We also propose the Ginibre point process as an alternative model.

\medskip
{\em Notation.}
Let $\calN$ denote the space of motion-invariant counting measures (point processes) on $\R^2$.
For $\calP\in\calN$, $V_{\calP}(x)$, $x\in\calP$, is the Voronoi cell
of $x$ and $\partial V_{\calP}(x)$ is its boundary. 
$\calP_o\triangleq (\calP\mid o\in\calP)$, and $\calP_o^!\triangleq\calP_o\setminus\{o\}$, where $o=(0,0)$.
$\E_o^!$ is the expectation
w.r.t.~the reduced Palm measure \cite{net:Haenggi12book}.
We let $U(B)$, $B\subset\R^2$, be a point chosen uniformly at random from $B$,
independently for different $B$ and
independent from everything else. 
Also we denote by $b(x,r)$ the disk of radius $r$ centered at $x\in\R^2$ and define
\[ S(v,r)\triangleq r^2\cos^{-1}(v/r)-v\sqrt{r^2-v^2},\quad 0\leq v\leq r, \]
which is the area of the disk segment $b((-v,0),r)\cap (\R^+\times \R)$.
As usual, $|B|$ is the area or length of $B$, and we define $\|B\|\triangleq\min\{\|x\|\colon x\in B\}$. 
Lastly, we use $\simeq$ to denote an approximation that becomes better
asymptotically.

\section{User Point Process of Type I}
This model is suitable for a fully loaded network, where each cell has an active user
in a given RB.
Its definition is based on a single point process only.
\begin{definition}[User point process of type I]
For $\calP\in\calN$, the user point process of type I is defined as
\[ \Phi \triangleq  \{x\in\calP\colon U(V_{\calP}(x)) \} .\]
\end{definition}
In this model, the user point process is obtained by a uniformly random displacement
of each BS $x\in\calP$ in its Voronoi cell.
By construction, $\Phi\in\calN$, with the same density as
$\calP$.
We first explore $\Phi$ when $\calP$ is a uniform PPP.

\subsection{Pair correlation function (pcf)}
\begin{figure}
 \centerline{\epsfig{file=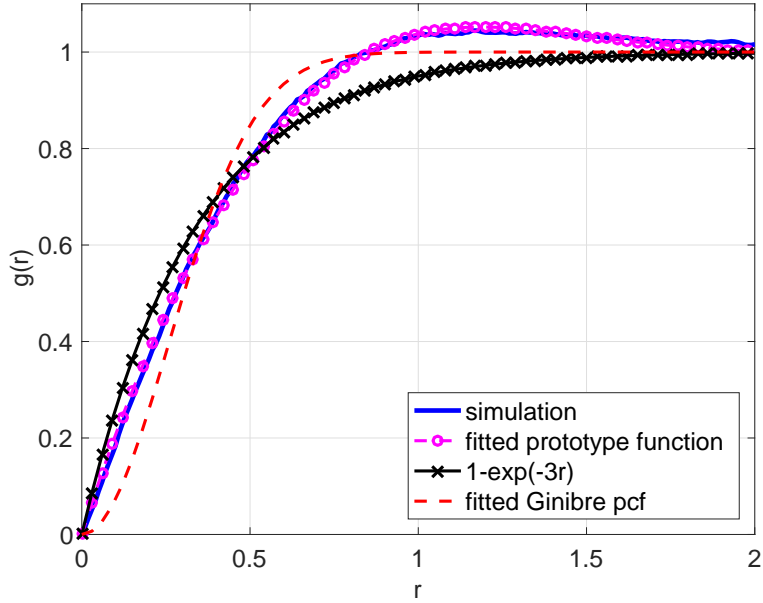,width=\figwidth}}
 \caption{Simulated pair correlation function $g(r)$ for the type I model and analytical approximations.
 The prototype function is \eqref{g_prototype},
 $1-e^{-3r}$ is the best fit for a simple exponential,
 and the fitted Ginibre pcf is \eqref{g_user_gin}.}
 \label{fig:g_user}
 \end{figure}

We consider the usual $K$ function by Ripley \cite[Def.~6.8]{net:Haenggi12book}, defined as
$K(r)\triangleq \E_o^! \Phi(b(o,r))$.
The pcf is
$g(r)\triangleq \frac1{2\pi r}K'(r)$.
We note that the model is scale-invariant, \ie, $g_\lambda(r)\equiv g_1(r\sqrt{\lambda})$, hence
we can focus just on $g_1=g$.
For $x\in\Phi$, we let $p_x\in\calP$ be the nucleus of the Voronoi cell that $x$ resides in, \ie,
the serving BS of user $x$.
\begin{lemma}
\label{lem:m1}
When $\calP$ is a uniform PPP of intensity $\lambda$,
\[ g(r)=\Theta(r),\quad r\to 0 .\]
\end{lemma}
\begin{IEEEproof}
We consider $\Phi_o$.
Let $D_V\triangleq \|\partial V(p_o)\|$ be the distance from $o$ to the nearest Voronoi cell boundary,
and let $y$ be the user located in the adjacent cell across that boundary.
The event that $o$ has a neighbor within distance $r$ is denoted by 
$E\triangleq\{\Phi_o\big(b(o,r)\cap V(p_y)\big)>0\}$. 
We write $\P(E)=\P(E_0)\P(E\mid E_0)$, where $E_0=\{D_V<r\}\supset E$.

Hence, for small $r$,
\vspace*{-3mm}
\begin{align*}
 K(r)&\sim \E(\P(E_0\mid \calP)\P(E\mid E_0,\calP)) ,\quad r\to 0\\
   &= \E\left(\frac{r|\partial V(p_o)|}{|V(p_o)|}\, \frac{S(D_V,r)}{|V(p_y)|} \right),
\end{align*}
where $r|\partial V(p_o)|$ is the area of the part of the Voronoi cell that is within
distance $r$ of its boundary. Similarly, $S(D_V,r)$ is the area of the part of the neighboring
 cell that $y$ needs to fall in.
We have $\P(E_0\mid\calP)=\Theta(r)$ while $\P(E\mid E_0, \calP)=\Theta(r^2)$ 
as the area of the disk segment grows with $r^2$. 
Since the expectation over $\calP$ 
does not change the exponent of $r$,
we have $K(r)=\Theta(r^3)$ and thus $g(r)=\Theta(r)$.
\end{IEEEproof}
\figref{fig:g_user} shows simulation results and approximations of $g(r)$.
With some simplifying assumptions, we can give an estimate for the slope of $g(r)$ at $r=0$,
\ie, the coefficient $a$ in $g(r)\sim ar$, $r\to 0$.
To estimate $a$, we assume that the Voronoi cells are square-shaped and of independent area $A=|V(p_o)|$
whose pdf follows the usual gamma approximation  \cite{net:Tanemura03}
\begin{equation}
  f_A(x)=\frac{\cc^\cc}{\Gamma(\cc)} x^{\cc-1} e^{-\cc x},
  \quad \cc\triangleq \frac72.
  \label{gamma_area}
\end{equation}
With the independence assumption, $K(r)\sim P_1 P_2$, where
\[ P_1\simeq \E\left(\frac{r|\partial V(p_o)|}{|V(p_o)|}\right)= \E\frac{4r}{\sqrt A} = \frac{32\sqrt{14}r}{15\sqrt{\pi}}\approx 4.50r. \]
For $P_2$, we note that
$D_V$ given $E_0=\{D_V< r\}$ is uniform in $[0,r)$ for small $r$. It follows that
\begin{align*}
P_2&\simeq \E S(D_V,r) \E(A^{-1}) 
   =\left( \frac{1}{r}\int_0^r S(v,r) \dd v\right) \E(A^{-1}) 
  =\frac{2r^2}{3} \E(A^{-1})=\frac{14}{15} r^2
 \end{align*}
 since $\E(1/A)=7/5$.
Hence $P_1P_2 \approx 4.20 r^3$, and, since $3\times 4.2(2\pi)\approx 2$, $g(r)\approx 2r$ for small $r$.

Since the simulated curve in \figref{fig:g_user} shows a maximum at $r=1.2$ before descending to $1$,
a natural candidate for an approximating analytical function is
\[ g_{\rm p}(r)\triangleq 1-e^{-ar}+br^2e^{-cr^2} \]
for $a,b,c>0$, henceforth referred to as the prototype function.
$a$ denotes the slope at $0$, as above.
Fitting yields $a=9/4$, $b=1/2$, $c=5/4$, hence a very good approximation is (see \figref{fig:g_user})
\begin{equation}
g(r) \approx 1-e^{-(9/4)r}+(1/2)r^2e^{-(5/4)r^2}.
\label{g_prototype}
\end{equation}
This shows that the relatively simple estimate above for the slope is only 12\% off.

\subsection{Base station/user pair correlation function}
Here we explore the point process of users as seen from a base station.
Formally, the point process of {\em interfering} users at the typical BS, assumed at $o$, is
\[ \Phi_{\rm BS} \triangleq \{x\in\calP_o^!\colon U(V_{\calP_o}(x))\}. \]
Since $\Phi_{\rm BS}$ is only isotropic but not stationary, we need
to specify that the pcf of interest is the one with respect to the origin,
hence we define $K_{\rm BS}(r)\triangleq \E\Phi_{\rm BS}(b(o,r))$ and, as before,
$\gbs(r)\triangleq K_{\rm BS}'(r)/(2\pi r)$.
Again we can focus on $\lambda=1$ due to scale-invariance.

\begin{figure}
 \centerline{\epsfig{file=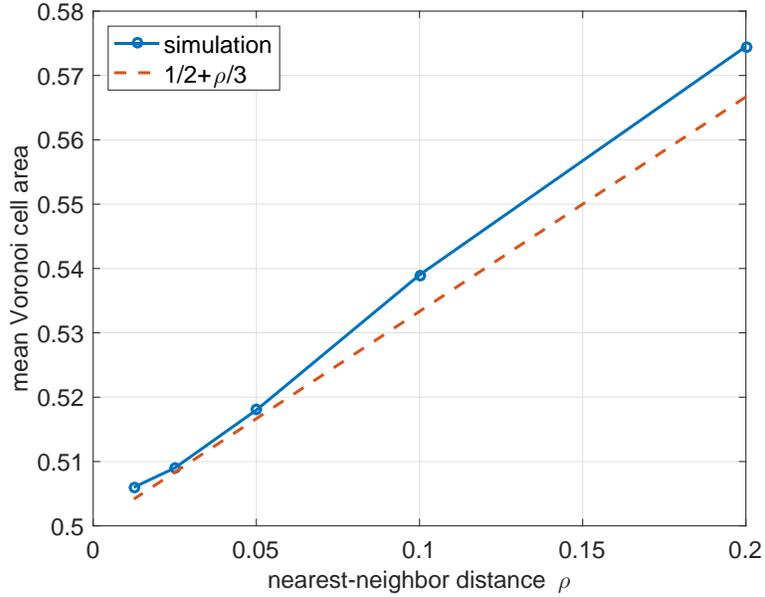,width=\figwidth}}
 \caption{Expected area of the Voronoi cells whose nucleus has a nearest neighbor within distance $\rho$.}
 \label{fig:small_cell_area}
 \end{figure}

\begin{lemma}
\label{lem:m1bs}
When $\calP$ is a uniform PPP of intensity $\lambda$,
\[ \gbs(r)=\Theta(r^2),\quad r\to 0 .\]
\end{lemma}
\begin{IEEEproof}
Let $R=\|\calP\|$, which is the distance of the BS at $o$ to its nearest neighbor $y$.
If $R\geq 2r$, then the nearest interfering user is farther than $r$.
If $R<2r$, then the user in BS $y$'s cell has probability $\E(S(R/2,r)/A_y)$ to be within distance $r$
of $o$, where $A_y$ is the area of the cell of $y$. 
We have
\begin{align*}
  \E\,S(R/2,r)&=\int_0^{2r} S(u/2,r) 2\pi u e^{-\pi u r^2} \dd u \sim \frac12 \pi^2 r^4, \quad r\to 0,
\end{align*}
since $1-e^{-\pi u^2}\sim \pi u^2$, $u\to 0$.
$A_y$ does not depend on $r$, so $K_{\rm BS}(r)=\Theta(r^4)$, $r\to 0$.
\end{IEEEproof}
As in the previous case, we can obtain an estimate for the pre-constant.
When approximating $\E(A_y^{-1})$, we need to take into account that 
$A_y$ is not the area of the typical cell but that
of a cell whose BS is very close to a boundary. Intuitively, the mean area of such a
skewed cell is about half the mean area of the typical cell. 
Detailed simulations confirm that this is indeed the case. In fact, if $\rho$ is the distance to
the nearest neighbor, $\E A(\rho)\approx 1/2+\rho/3$, for small $\rho$,
see \figref{fig:small_cell_area}. Hence, assuming the shape of the cell is the
same as that of the typical cell, we can use \eqref{gamma_area} with an adjusted mean
and obtain $\E(A_y^{-1})\approx 14/5$.

\begin{figure}
 \centerline{\epsfig{file=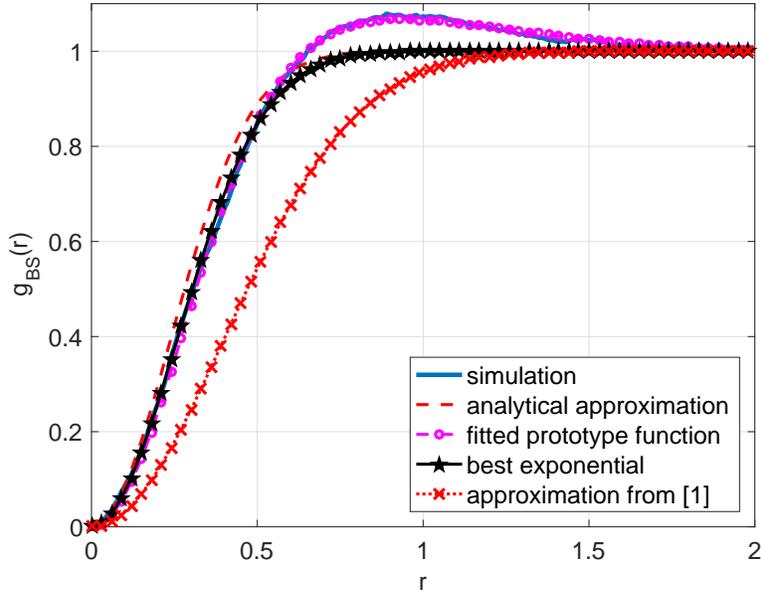,width=\figwidth}}
 \caption{Simulated base station/user pair correlation function $\gbs(r)$ for the type I model and approximations.
The ``analytical approximation" is
 \eqref{gbs_anal}, the prototype function is \eqref{gbs_prototype}, and the best exponential is \eqref{gbs_exp}.
 The last curve, for comparison, is the approximation $1-e^{-\pi r^2}$ from \cite{net:Singh15twc}.}
 \label{fig:g_bs}
 \end{figure}

Combining the results for $\E S(R/2,r)$ and $\E(A_y^{-1})$, we obtain
$K_{\rm BS}(r) \approx \frac{7}{5}\pi^2 r^4$ and thus, for small $r$,
$\gbs(r) \approx \frac{14}{5}\pi r^2\approx 8.8 r^2$.
Hence we expect 
\begin{equation}
   \gbs(r)\simeq 1-e^{-14\pi r^2 /5} 
 \label{gbs_anal}
\end{equation}
to be near-exact for small $r$ and a decent approximation for all $r$.
In view of Lemma \ref{lem:m1bs} and the simulation result in \figref{fig:g_bs}, the natural prototype function here is of the form
\begin{equation}
   g_{\rm p,BS}(r)\triangleq 1-e^{-ar^2}+br^2e^{-cr^2} ,\quad a,b,c>0.
  \label{gbs_prototype}
\end{equation}
Fitting yields $a=13/2$, $b=2/7$, and $c=13/9$.
The best exponential fit, also obtained numerically, is
\begin{equation}
   \gbs(r)  \approx 1-e^{-12\pi r^2/5}. 
 \label{gbs_exp}
\end{equation}
 \subsection{Mean interference} 
The mean interference at the typical user from other users is 
$\E_o^!(I)=2\pi \int_0^\infty g(r) r^{1-\alpha} \dd r$ \cite{net:Haenggi11cl}.
If $g(r)=\Theta(r)$, a necessary condition for the mean interference to be finite is
$\alpha<3$, since $\int_0^\epsilon r^{2-\alpha}\dd r$ for $\epsilon>0$ is finite only if $\alpha<3$.
Hence Lemma \ref{lem:m1} implies that the mean interference is finite only for
$2<\alpha<3$.
For the mean interference at a base station (uplink), $\gbs(r)=\Theta(r^2)$, $r\to 0$,
 thus the range of $\alpha$ that results in finite mean interference is
$2<\alpha<4$.
In this case, for $\gbs(r)=1-e^{-a\pi r^2}$,
$\E(I_{\rm BS})=\pi^{\alpha/2} a^{\alpha/2-1}\Gamma(1-\alpha/2)$.
Therefore, for $3.6<\alpha<4$, if the approximation in \cite{net:Singh15twc} was used for
heavily loaded networks, it would underestimate
the mean interference by more than a factor $2$.

\subsection{Approximation with PPP}
Here we consider the user point process $\Phi$ of general intensity $\lambda$.
We would like to find a PPP $\ppp$ such that for all $f\colon \R^2\mapsto \R^+$,
\begin{equation}
   \E_o^!\sum_{x\in\Phi} f(x)\equiv \E \sum_{x\in\ppp} f(x). 
   \label{equiv}
\end{equation}
Expanding both the left and right sides, using Campbell's theorem on the right and its Palm version
(where the reduced second factorial moment measure replaces the intensity measure
\cite{net:Haenggi12book}) on the left, the identity can be formulated as
\[ \lambda\int_{\R^2} f(x) g(\|x\|\sqrt\lambda) \dd x\equiv \int_{\R^2}f(x)\lppp(\|x\|)\dd x ,\]
which is satisfied if we set the intensity function to
\begin{equation}
   \lppp(r):=\lambda g(r\sqrt{\lambda}).
   \label{lppp}
\end{equation}
The same idea was put forth in \cite[Assumption 1]{net:Singh15twc}.
In principle any point process with intensity function $\lppp$ could be used,
but the PPP is a natural choice due its unparalleled tractability.
Accordingly, for the user point process, 
\[ \lppp(r):=\lambda\big(1-e^{-(9/4)\sqrt\lambda r}+(1/2) \lambda r^2 e^{-(5/4)\lambda r^2}\big) ,\]
or, for a coarser but simpler approximation,
\begin{equation}
   \lppp(r):=\lambda\big(1-e^{-3 \sqrt\lambda r} \big) .
   \label{lppp_coarse}
\end{equation}

From the perspective of the typical BS, we replace $g$ by $\gbs$ in \eqref{lppp} and set
\begin{align}
   \lppp^{\rm BS}(r)&:=\lambda\big(1-e^{-(13/2)\lambda r^2}+(2/7) \lambda r^2 e^{-(13/9)\lambda r^2}\big) \nonumber\\
   \intertext{or}
   \lppp^{\rm BS}(r)&:=\lambda\big(1-e^{-(12/5)\pi\lambda r^2}) .
   \label{lppp_bs_coarse}
\end{align}

\subsection{Approximation with Ginibre point process}
The $\beta$-Ginibre point process\footnote{This process is obtained by
independent thinning of a Ginibre process of intensity $1$ with retention
probability $\beta$ and subsequent rescaling to preserve the intensity.}
 \cite{net:Deng15twc} has the pcf 
$g_{\rm gin}(r)= 1-e^{-\pi r^2/\beta}$, which is strikingly
similar to that of the interfering users at a base station.
Hence the $\beta$-Ginibre process with $\beta=5/12$, denoted by $\Phi_{\rm gin}$, has $g_{\rm gin}(r) \approx \gbs(r)$.
Since the Ginibre process also models the repulsion between users, not just
the repulsion between the BS and the users, it is a more accurate approximation
than the PPP, albeit at reduced tractability. 
It also provides an efficient
way of simulating $\Phi_{\rm BS}$ especially if only the distances matter, since
the moduli of the points of the Ginibre process are equal in distribution to a set of
independent gamma random variables \cite[Sec.~III.A]{net:Deng15twc}.

If the Ginibre process was used to approximate the user process $\Phi$ (well
aware that the slope at $0$ does not match), the best fit is obtained for
\begin{equation}
g_{\rm gin}(r)=1-e^{-\pi (12/5) r^2},
\label{g_user_gin}
\end{equation}
which, remarkably, is exactly the same approximation as for $\Phi_{\rm BS}$.
As seen in \figref{fig:g_user}, the gap to the true curve here is bigger, but
the $\beta$-Ginibre process still is a decent substitute for the user
point process of type I.
The Ginibre approximation also shows that independent thinning does not
affect the pcfs $g$ and $\gbs$.

\subsection{Nearest-neighbor distance distribution}
Let $D=\|\Phi_o^!\|$.
From the resemblance to the Ginibre process and \cite[Prop.~1]{net:Deng15twc}, an educated guess for the
nearest-neighbor distance distribution $f_D$ is that $D^2$ follows a gamma
distribution with parameters $2$ and $1/5$, since simulation results indicate $\E(D^2)\approx 2/5$.
It follows that
\begin{equation}
   f_D(r)\approx 50 r^3 e^{-5r^2} ,
 \end{equation}
which results in a mean $\E(D)=0.594$, in
perfect agreement with simulation results. The same approximation can be used also for
the distribution of the nearest interfering user distance $\|\Phi_{\rm BS}\|$.

\subsection{User-base station distance distribution}
Since the typical user resides in the typical cell, the distance $R=\|p_o\|$ to its serving BS
is not the standard Rayleigh distribution with mean $1/2$ (for $\lambda=1$), but a similarly-shaped
distribution with (empirical) mean $0.439$, corresponding to a correction factor $13/10$
in the density. Hence a very good approximation for general $\lambda$ is
\begin{equation}
   f_R(r)\approx 2(13/10)\pi r\lambda e^{-(13/10)\lambda\pi r^2} .
\end{equation}

\subsection{Users in lattice processes}
\begin{lemma}
For the stationary square lattice $\calP=U([0,1]^2)+\mathbb{Z}^2$, the pcf
 of the user process $\Phi$ satisfies
\[ g(r)\sim \frac{8}{2\pi} r,\quad r\to 0.\]
\end{lemma}
\begin{IEEEproof}
Following the same procedure as for the Poisson case, we can determine the
behavior of $g(r)$ for small $r$.
In this case,
$K(r)\sim 4r \E S(D_V,r) = \frac{8}{3} r^3$, $r\to 0$.
\end{IEEEproof}
Due to the concavity of the pcf, $g(r)\lesssim \min\{8r/(2\pi),1\}$ is a good upper bound,
and $g(r)\gtrsim 1-e^{-8r/(2\pi)}$ is the corresponding lower bound. Both
are asymptotically exact.

\section{User Point Process of Type II}
\subsection{Definition}
\begin{definition}[User point process of type II]
For independent $\Phi_0,\calP\in\calN$, the user point process of type II is defined as
 \[ \Phi\triangleq\{x\in\calP\colon U(V_\calP(x)\cap\Phi_0)\}. \]
\end{definition}
Again $\calP$ models the BSs.
Here the user in each cell $V_\calP(x)$ is chosen only from the countable set $V_\calP(x)\cap\Phi_0$,
where $\Phi_0$ can be viewed as the entire user population, while $\Phi$ are those served in
the RB under consideration. 

\subsection{Properties for the Poisson/Poisson case}
The simplest model here is that $\Phi_0$ and $\calP$ are independent uniform PPPs of 
intensities $\lambda_0$ and $\lambda_{\calP}$. In this case, a fraction $\nu\triangleq\E(e^{-\lambda_0 A})$
of cells is vacant, where $A$ is the area of the typical Voronoi cell of $\calP$, and the density of the
user process $\Phi$ is
$\lambda=\lambda_{\calP}(1-\nu)$.
Using the approximation \eqref{gamma_area}, 
\begin{equation}
  \nu\approx (\cc\lambda_{\calP})^\cc(\cc\lambda_{\calP}+\lambda_0)^{-\cc}=(\cc/(\cc+\eta))^\cc ,
  \label{nu}
\end{equation}
where $\eta=\lambda_0/\lambda_{\calP}$.
We distinguish three regimes:
\subsubsection{$\eta\ll 1$ (low user density)}
Here, $\nu\approx 1$, and most cells are empty, \ie, $\lambda\approx\lambda_0\ll\lambda_{\calP}$. Those
that are occupied are, on average, larger than the typical cell, \ie, they are Crofton
cells (or 0-cells)  \cite{net:Mecke99pattern}, which have a mean area
$\E(A_0)=\E(A^2)/\E(A)$. For the PPP, $\E(A_0)\approx (9/7)\lambda_{\calP}^{-1}$.
The link distance $R$ follows the standard Rayleigh distribution
$f_R(r)=2\lambda_{\calP}\pi r e^{-\lambda_{\calP}\pi r^2}$ with mean $1/(2\sqrt{\lambda_{\calP}})$,
and the BS/user pcf $\gbs(r)=1-e^{-\lambda_{\calP}\pi r^2}$ from \cite{net:Singh15twc} becomes accurate
since $\Phi\approx \Phi_0$ is essentially still a PPP.

\subsubsection{$\eta\gg 1$ (high user density)} In this regime, the model behaves as
the model of type I (essentially no cells are empty).

\subsubsection{$\eta\approx 1$ (intermediate regime)} For $\eta\in [1/2,2]$, $60\%$-$20\%$ of the cells
are empty, and the mean link distance falls in the regime $[0.46,0.48]/\sqrt{\lambda_{\calP}}$. The mean
area of the occupied cell is in the range $[1.1,1.2]/\lambda_{\calP}$.

\begin{figure}
 \centerline{\epsfig{file=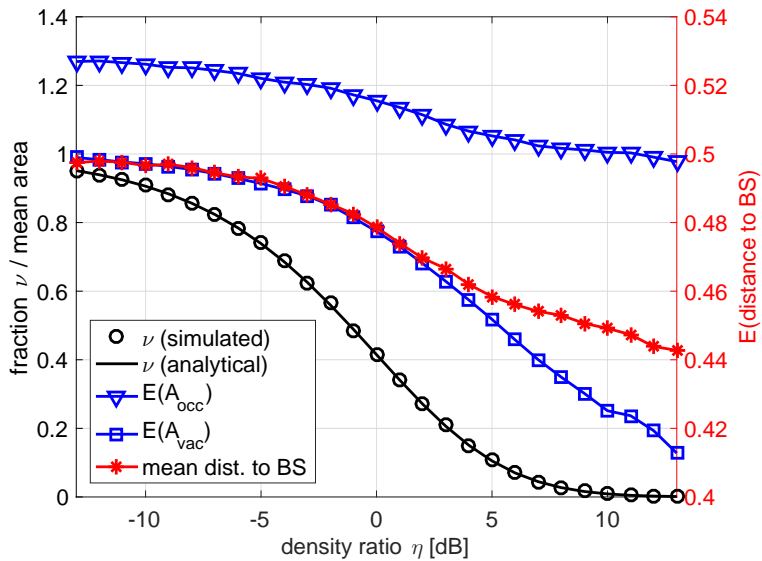,width=\figwidth}}
 \caption{Properties of the user model of type II as a function of
 the density ratio $\eta$ (dB) for $\lambda_{\calP}=1$. The left $y$ axis applies to the fraction of vacant cells given in \eqref{nu}
 and the mean areas of occupied and vacant cells $\E(A_{\rm occ})$ and $\E(A_{\rm vac})$. The right axis
 applies to the mean distance $\E(R)$ between user and serving BS.
 Interestingly, $\E(A_{\rm occ})/\E(R)^2\approx 5$ for all values of $\eta$.}
 \label{fig:model2}
 \end{figure}

\figref{fig:model2} summarizes and illustrates these properties. Note that for the 
user-BS distance, the right $y$ axis applies.
In general, since the average cell size is $\lambda_{\calP}^{-1}$, the mean areas of
vacant and occupied cells are related as $1/\lambda_{\calP} = \nu\E(A_{\rm vac})+(1-\nu)\E(A_{\rm occ})$.

\section{Conclusions}
Even in cellular networks whose BSs are modeled as a PPP, active users that are served in the same RB do not form a
PPP but a soft-core process. Indeed, in the case where BSs are heavily loaded, the user point process
shows interesting connections to the Ginibre point process.
This case is naturally represented using models of type I but can also be
achieved with models of type II if the ratio parameter $\eta\gg 1$.
Lightly loaded cellular networks can be modeled by  
applying independent thinning to the type I model, which does not affect the
pcfs, or using the type II model whose pcfs depend on the density ratio $\eta$.


 \end{document}